# Phase transitions of dense neutron matter with generalized Skyrme interaction to superfluid states with triplet pairing in strong magnetic field

A N Tarasov

Institute for Theoretical Physics, NSC "Kharkov Institute of Physics and Technology", Kharkov 61108, Ukraine

E-mail: antarasov@kipt.kharkov.ua

**Abstract**. A generalized non-relativistic Fermi-liquid approach was used to find analytical formulas for temperatures $T_{c1}(n,H)$ and $T_{c2}(n,H)$ (which are functions nonlinear of density $n$ and linear of magnetic field $H$) of phase transitions in spatially uniform dense pure neutron matter from normal to superfluid states with spin-triplet p-wave pairing (similar to anisotropic superfluid phases $^3$He-A$_1$ and $^3$He-A$_2$) in steady and homogeneous strong magnetic field (but $|\mu_n|H \ll E_c < \varepsilon_F(n)$, where $\mu_n$ is the magnetic dipole moment of a neutron, $E_c$ is the cutoff energy and $\varepsilon_F(n)$ is the Fermi energy in neutron matter). General formulas for $T_{c1,2}(n,H)$ (valid for arbitrary parameterization of the effective Skyrme interaction in neutron matter) are specified here for generalized BSk18 parameterization of the Skyrme forces (with additional terms dependent on density $n$) on the interval $0.3 \cdot n_0 < n < n_c(\text{BSk18}) \approx 2.7952 \cdot n_0$, where $n_0 = 0.17 \text{ fm}^{-3}$ is nuclear density and at critical density $n_c(\text{BSk18})$ triplet superfluidity disappears, $T_{c0}(n_c, H=0)=0$. Expressions for phase transition temperatures $T_{c0}(n) < 0.09$ MeV (at $E_c = 10$ MeV) and $T_{c1,2}(n,H)$ are realistic non-monotone functions of density $n$ for BSk18 parameterization of the Skyrme forces (contrary to their monotone increase for all previous BSk parameterizations). Phase transitions to superfluid states of such type might occur in liquid outer core of magnetars (strongly magnetized neutron stars).

## 1. Introduction

Note, that neutrons are presumably the main constituent of neutron stars (NSs). In view of the fact that there is no decisive evidence that only $^3$P$_2$ pairing of neutrons [1, 2] occurs in outer core of NSs (it is not yet clear how n-n interaction in vacuum is modified in medium of NS at supra-saturation densities) and for lack of consensus in the calculated magnitude of neutron $^3$P$_2$ pairing gap at high densities [3], [4], we consider here another but somewhat simplified problem. Namely, in this work we shall focus our study on spin-triplet superfluid pure neutron matter (SPNM) which is spatially homogeneous (and, as a consequence, without spin-orbit coupling) at sub- and supra-nuclear densities from the interval $0.3 \cdot n_0 < n < 2.5 \cdot n_0$ (where $n_0 = 0.17 \text{ fm}^{-3}$ is number density of nuclear matter) with p-wave pairing

(similar to anisotropic superfluid phases $^3$He-A$_1$ and $^3$He-A$_2$, see [1]) in strong spatially uniform magnetic field H. Upper limit for magnetic fields studied here may be of the order $10^{17}$ G or even more (as inside cores of magnetars, i.e., strongly magnetized neutron stars [5, 6]). The so-called generalized BSk18 parameterization of Skyrme forces with additional terms depending on density (details can be found in [7]) are used here as interactions in SPNM. The general consensus is that relativistic effects are small up to $(2 \div 3) \cdot n_0$ and the Skyrme interaction is well justified for the description of nuclear matter consisting of nucleons (no strange baryons or mesons) (see, e.g., [8] and references therein). That is why we shall use the generalized non-relativistic Fermi-liquid approach [9] which is the extension of the Landau's theory of normal Fermi-liquid to superfluid Fermi-liquids.

Obtained here expressions for phase transition temperatures of dense NM to spin-triplet superfluid states $T_{c0}(n, H=0)$ and $T_{c1,2}(n, H)$ are realistic non-monotone functions of density $n$ for BSk18 parameterization of the Skyrme forces (contrary to their monotone increase for all previous BSk and other conventional parameterizations [12]). This work is a continuation of our previous investigations [10] and [11] (published in Proceedings of LT23 and LT24; see also more detailed article [12] and references therein) which were devoted to theoretical description of dense SPNM with conventional Skyrme forces and with the same type of anisotropic spin-triplet pairing in strong magnetic fields.

## 2. General equations for the order parameter and effective magnetic field for SPNM with the Skyrme forces and triplet pairing

Using general formulas for anomalous and normal distribution functions of quasiparticles [13, 14] (neutrons) for SPNM in magnetic field we have derived a set of integral equations for $\xi(p)$ and $\Delta_\uparrow^{A_2}$, $\Delta_\downarrow^{A_2}$. Namely, for the effective magnetic field (EMF) in SPNM we have $\vec{\xi}(\mathbf{p}) = \xi(p)\mathbf{H}/H \equiv -\mu_n \mathbf{H}_{\text{eff}}(p)$ ($\mu_n \approx -0.60308 \cdot 10^{-17}$ MeV/G is the magnetic dipole moment of neutron [15]) and for $\xi(p)$ we obtain the equation [10, 12]:

$$\xi(p) = -\mu_n H + (r + sp^2)K_2(\xi) + sK_4(\xi). \tag{1}$$

Here $r = t'_0 + (t'_3/6)n^\alpha$ and $s = (t'_1 - t'_2)/(4\hbar^2)$, $n \equiv yn_0$ is density of neutron matter; $t'_0 = t_0 \cdot (1 - x_0)$, $t'_1 = t_1 \cdot (1 - x_1)$, $t'_2 = t_2 \cdot (1 + x_2)$, $t'_3 = t_3 \cdot (1 - x_3)$ and $1/6 \leq \alpha \leq 1/3$ are parameters of the Skyrme interaction (cf. [8], [12]). The functionals $K_\beta(\xi)$ ($\beta = 2, 4$) in equation (1) have the form:

$$K_\beta(\xi) = \frac{1}{8\pi^2 \hbar^3} \int_{p_{\min}}^{p_{\max}} dq\, q^\beta \int_0^1 dx\, \kappa(q, x), \tag{2}$$

where

$$\kappa(q, x) = \frac{z(q) + \xi(q)}{E_+(q, x^2)} \tanh\left(\frac{E_+(q, x^2)}{2T}\right) - \frac{z(q) - \xi(q)}{E_-(q, x^2)} \tanh\left(\frac{E_-(q, x^2)}{2T}\right), \tag{3}$$

$$E_\pm^2 = q^2 \Delta_{\uparrow(\downarrow)}^2 (1 - x^2) + (z(q) \pm \xi(q))^2, \tag{4}$$

$z(q) = q^2/2m_n^* - \mu$ ($m_n^*$ is the effective mass of neutron dependent on density [11, 12], $\mu$ is the chemical potential). We have taken into account that for SPNM with pairing of the $^3$He-A$_2$ type the order parameter (OP) can be written as $\Delta_{\uparrow(\downarrow)}^{A_2}(T, \xi, q) = q\Delta_{\uparrow(\downarrow)}(T, \xi)$, where functions $\Delta_{\uparrow(\downarrow)}(T, \xi)$ obey the following equations [10, 12]:

$$\Delta_{\uparrow(\downarrow)}(T, \xi) = -\Delta_{\uparrow(\downarrow)}(T, \xi) \frac{c_3}{8\pi^2 \hbar^3} \int_{p_{\min}}^{p_{\max}} dq\, q^4 \int_0^1 dx(1 - x^2) \frac{\tanh(E_\pm(q, x^2)/2T)}{E_\pm(q, x^2)} \tag{5}$$

($p_{\max} > p_F > p_{\min}$, $(p_{\max} - p_{\min})/p_F \ll 1$, $p_F$ is the Fermi momentum). Here $c_3 \equiv t_2(1 + x_2)/\hbar^2 < 0$ is coupling constant leading to spin-triplet p-wave pairing of neutrons, which is expressed through the parameters $t_2$ and $x_2$ of the Skyrme interaction. Note that we consider here a traditional model of neutron Cooper pairing in a thin shell symmetric about the Fermi sphere, $p_{max} - p_F = p_F - p_{min}$.

This set of nonlinear integral equations (1) and (5) for the EMF and OP give us the possibility to describe thermodynamics of superfluid non-unitary phases of $^3$He-A$_{1,2}$ type in dense SPNM with spin-triplet p-wave pairing in static uniform high magnetic field at arbitrary temperatures from the interval $0 \leq T \leq T_c(H)$. In general case these equations can't be solved analytically and it is necessary to use numerical methods for their solving. But we can solve equations (1), (5) using analytical methods in the limiting case, when the temperature ($T \lesssim T_{c0}$) is near the PT temperature $T_{c0}(n)$ of dense NM to superfluid state, and it is the theme of the next section.

### 3. Solutions of equations for the OP and EMF for dense neutron matter with generalized Skyrme forces near $T_{c0}$

Note, that Fermi energy in neutron matter (NM) for the BSk18 parameterization [7] of the Skyrme forces has the following form

$$\varepsilon_{F,\,BSk18}(y) \approx y^{2/3} \cdot (1 + 0.253920 \cdot y) \cdot 60.902 \text{ (MeV)}, \tag{6}$$

where $y \equiv n/n_0$ is the density of SPNM reduced to dimensionless form.

In zero magnetic field $H = 0$ equations (5) are reduced to single equation for the energy gap in SPNM from which we have obtained as a result the following expression for phase transition (PT) temperature of dense NM to SPNM with anisotropic spin-triplet pairing of $^3$He-A type studied here:

$$T_{c0;\,BSk18}(E_c; y) \approx 1.14055 \cdot E_c \cdot \exp\left[\frac{5.0552 \cdot 10^{-5} \cdot E_c^2}{y^{4/3} \cdot (1 + 0.253920 \cdot y)^2} + \frac{4.26 \cdot 10^{-10} \cdot E_c^4}{y^{8/3} \cdot (1 + 0.253920 \cdot y)^4}\right]$$

$$\times \exp\left[\frac{1 + 0.253920 \cdot y}{y \cdot (0.139400 \cdot y - 0.389652)}\right] \text{ (MeV)}, \tag{7}$$

(here $E_c < \varepsilon_{F,BSk18}(y)$ is the cutoff energy). Now we write down the final formulae (obtained from the equations (1) and (5)) for the PT temperatures $T_{c1;\,BSk18}(E_c; H, y)$ and $T_{c2;\,BSk18}(E_c; H, y)$ of dense NM in strong magnetic field $H$ to superfluid states of $^3$He-A$_1$ and then to $^3$He-A$_2$ type respectively:

$$T_{c1,2;\,BSk18}(E_c; H, y) \approx T_{c0;\,BSk18}(E_c; y) \cdot \left\{1 \mp \frac{Z \cdot 0.60308}{\varepsilon_{F,\,BSk18}(y) \cdot I_{0,\,BSk18}(E_c; y)}\left[A_{c0,\,BSk18}(E_c; y) \cdot I_{A,\,BSk18}(E_c; y) + \right.\right.$$

$$\left.\left. + B_{c0,\,BSk18}(E_c; y) \cdot I_{B,\,BSk18}(E_c; y)\right]\right\} \text{ (MeV)}, \tag{8}$$

where upper sign "−" corresponds to $T_{c1;\,BSk18}(E_c; H, y)$ and lower sign "+" - to $T_{c2;\,BSk18}(E_c; H, y)$. We have introduced here variable Z defined as $H \equiv Z \cdot 10^{17}$ (G), so that $|\mu_n| H \approx Z \cdot 0.60308$ (MeV). Note also that the functions $A_{c0,\,BSk18}(E_c; y)$, $B_{c0,\,BSk18}(E_c; y)$ and integrals $I_{0,\,BSk18}(E_c; y)$, $I_{A,\,BSk18}(E_c; y)$, $I_{B,\,BSk18}(E_c; y)$ in (8) are of the same form as general functions $A(a; y, t_{c0})$, $B(a; y, t_{c0})$ and integrals $I_0$, $I_A$, $I_B$ in [12] (see also [11]) (which are valid for all Skyrme parameterizations) but with cutoff parameter $a \equiv E_c / \varepsilon_{F,BSk18}(y) < 1$ and with generalized BSk18 Skyrme parameters from [7].

Now, for the definiteness, we select cutoff energy $E_c = 10$ (MeV) and plot figures for the PT temperatures (7) and (8) (at $H = 10^{17}$ (G), i.e. at $Z = 1$, and on the interval $0 < H < 10^{17}$ (G)) of NM at sub- and suprasaturation densities on the interval $0.3 \cdot n_0 < n < 2.5 \cdot n_0 < n_c$ (BSk18) $\approx 2.7952 \cdot n_0$ (where $E_c < \varepsilon_{F,BSk18}(y)$ and at critical density $n_c$ (BSk18) triplet superfluidity disappears, $T_{c0}(n_c, H=0)=0$).

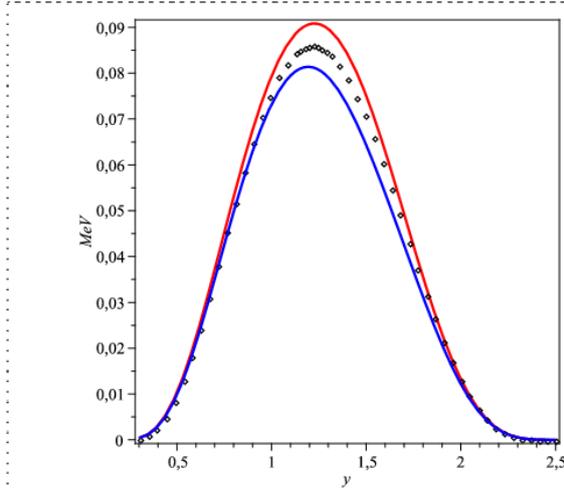 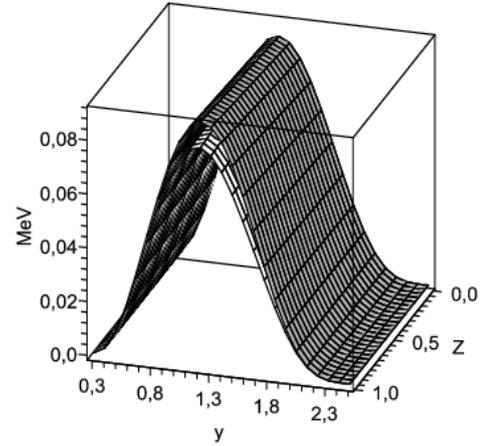

| **Figure 1.** PT temperatures (7) and (8) for BSk18 at $E_c=10$ (MeV) and $H=10^{17}$ (G) (Z=1): $T_{c1;BSk18}(10;Z=1,y)$ -upper curve; $T_{c0;BSk18}(10;y)$ - points; $T_{c2;BSk18}(10;Z=1,y)$ -lower curve. | **Figure 2.** PT temperatures (8) at $E_c=10$ (MeV) and $0<H<10^{17}$ (G): $T_{c1;BSk18}(10;Z,y)>T_{c2;BSk18}(10;Z,y)$. |

## 4. Conclusion

We have obtained that PT temperature $T_{c0;BSk18}(E_c;y)$ (see (7)) of dense NM (in zero magnetic field) to superfluid state with anisotropic p-wave pairing of $^3$He-A type and with generalized BSk18 Skyrme interaction [7] depends on density in non-monotone way (exhibits a bell-shaped density profile, see figure 1). Such behavior of $T_{c0;BSk18}(E_c;y)$ is in qualitative agreement with [16, 17] and is of the same order in magnitude at $E_c=10$ (MeV). It follows from (7) that $^3$He-A type of triplet superfluidity disappears in NM when density $n$ tends to $n_c(\text{BSk18}) \approx 2.7952 \cdot n_0$. Our main results are expressions (8) (and figures 1 and 2) for PT temperatures $T_{c1,2;BSk18}(E_c;H,y)$ linear in strong magnetic fields (which may approach to $10^{17}$ G or even more as in liquid outer cores of magnetars [5, 6]).